\documentclass[journal=jacs, manuscript=article, layout=twocolumn]{achemso}
\setkeys{acs}{maxauthors = 10, etalmode=truncate}
\setkeys{acs}{keywords = true}

\usepackage{etoolbox}

\usepackage[version=3]{mhchem}
\usepackage[T1]{fontenc}
\usepackage{multicol}
\usepackage{multirow}
\usepackage{mathrsfs}
\usepackage{amsmath}
\usepackage{relsize}
\usepackage{siunitx}
\usepackage{hyperref}
\usepackage{float}
\usepackage{cuted}
\usepackage{threeparttable}
\usepackage{pdfpages}

\makeatletter
\let\l@addto@macro\relax
\makeatother
\usepackage[fontsize=11pt]{scrextend}

\author{Allison R. Pessoa}
\affiliation[IFPE]
{Federal Institute of Education, Science and Technology of Pernambuco, Recife (50740-545), Pernambuco, Brazil}
\alsoaffiliation[UFPE]
{Department of Physics, Universidade Federal de Pernambuco, Recife (50740-540), Pernambuco, Brazil}
\email{allison.pessoa@ufpe.br}

\author{Leonardo de S. Menezes}
\affiliation[LMU]
{Chair in Hybrid Nanosystems, Faculty of Physics, Ludwig-Maximilians-Universität München, München (80539) Bavaria, Germany}
\alsoaffiliation[UFPE]
{Department of Physics, Universidade Federal de Pernambuco, Recife (50740-540), Pernambuco, Brazil}

\author{Anderson M. Amaral}
\affiliation[UFPE]
{Department of Physics, Universidade Federal de Pernambuco, Recife (50740-540), Pernambuco, Brazil}

\title{Addressing Discrepancies Between Theory and Experiments in Boltzmann Luminescence \\Thermometry with Ln$^{3+}$ Ions}

\keywords{Lanthanide Ions; Boltzmann thermometry; Primary thermometry; Optical thermometer; Stark sublevels
\vspace{0.5cm}}

\let\oldmaketitle\maketitle
\let\maketitle\relax

\begin{document}
\twocolumn[
\begin{@twocolumnfalse}
\oldmaketitle
\begin{abstract}

Trivalent lanthanide ion-doped nanoparticles are widely employed as nanoscale thermometers, driving rapid advancements in real-world applications. When the Luminescence Intensity Ratio (LIR) technique is used, these thermometric systems typically require a calibration process to obtain macroscopic calibration parameters. However, despite extensive studies from various research groups, significant discrepancies are observed among the reported values, even for identical Ln$^{3+}$-host systems under similar experimental conditions. Also, in many cases, the obtained calibration parameters substantially differ from their microscopic counterparts, which is commonly ignored in the literature. This study addresses some sources for these inconsistencies by providing fundamental theoretical insights into the measurement process. We demonstrate that the thermalization of the electronic population within the Stark sublevels of a given manifold plays a crucial role in the LIR's temperature dependence and consequently in measuring the macroscopic parameters. As a result, attempts to construct primary thermometers without prior calibration can result in temperature measurement errors exceeding 20 K. Additionally, we show that pathways disrupting Boltzmann thermalization, influenced by experimental conditions, also affect the evaluation of the macroscopic quantities. These findings contribute to a more robust theoretical framework for interpreting and understanding ratiometric Boltzmann luminescence thermometry experiments, also paving the way for developing more accurate and reliable primary thermometers.
\end{abstract}
\end{@twocolumnfalse}
]

\section{Introduction}

With the rapid advancement of nanoscale technologies, measuring temperature at the nanoscale has become essential for controlling and investigating physical and chemical processes at this level \cite{Brites_2012}. In this context, trivalent lanthanide ions (Ln$^{3+}$)-doped nanoparticles have been extensively used over the past decades to provide accurate nanoscale temperature measurements \cite{Puccini_2024, Dacanin_2023, Wang_2022, Brites_2019}. These ions interact with the host matrix, enabling optical temperature measurements through their luminescence or absorption spectra. Since the electronic transitions of Ln$^{3+}$ ions in the visible and near-infrared ranges occur within the $4f$ subshell, they give origin to sharp absorption and emission lines, long-lived excited states, and remarkable resistance to photobleaching \cite{Brites_2019, Reisfeld_2015}. These unique photophysical properties make Ln$^{3+}$ ions exceptional candidates for optical temperature sensing.

Several methods can retrieve the temperature information from optical absorption or emission signals, such as analyzing wavelength shifts or the luminescence intensity of specific electronic transitions \cite{Brites_2019}. Among these techniques, the Luminescence Intensity Ratio (LIR) method is the most widely used. It relies on the ratio between two or more luminescence bands' intensities, addressing limitations associated with single-intensity-based measurements, such as variations in excitation intensity or fluctuations in probe concentration. When two energy levels, referred to as Thermally Coupled Levels (TCLs), are close enough in energy to effectively allow non-radiative electronic exchange via phonon-mediated transitions, their relative electronic populations tend to follow the Boltzmann distribution \cite{Suta_Meijerink_2020}. This relationship yields an exponential temperature dependence of the LIR, expressed as $C \cdot \exp(-\Delta E / k_B T)$, where $k_B$ is the Boltzmann constant, $T$ is the absolute temperature, and $C$ and $\Delta E$ are often considered as calibration parameters, labeled as $C_\text{eff}$ and $\Delta E_\text{eff}$, determined through fitting LIR values at known temperatures, measured with the aid of external thermometers. The exponential temperature dependence forms the basis of what is known as Boltzmann luminescence thermometry.

Despite substantial experimental progress in practical applications, challenges remain in comprehensively understanding the relation between the microscopic properties of the Ln$^{3+}$-host system and the Boltzmann calibration parameters. For instance, Table \ref{tab:nayf4_comparison} presents data from 8 different works employing NaYF$_4$: 20\% Yb$^{3+}$ / 2\% Er$^{3+}$ as Boltzmann thermometers, exemplifying the high variability usually found in the literature. Additionally, $\Delta E_\text{eff}$ is usually considered in the literature as equal to the energy difference between the barycenters of the TCLs' luminescence spectral bands, here referred to as $\Delta E_\text{bary}$. However, Wang \textit{et al.} compared the quantities $\Delta E_\text{eff}$ and $\Delta E_\text{bary}$ from 35 studies employing Er$^{3+}$-based thermometers in different hosts \cite{Wang_2015} and found differences up to 90\% between these quantities, depending on the host matrix. In this work, we provide a solid theoretical explanation of how these quantities are interconnected. 

Our present work also shows that $\Delta E_\text{eff}$ depends on the photophysical dynamics of the Ln$^{3+}$-host system, which may include possible cross-relaxation effects \cite{Suta_Miroslav_2020}, excited state absorption \cite{Pickel_2018}, interaction with surrounding ligands \cite{Galindo_2021, GalindoCorr_2021}, among others. These influences could account for the observed discrepancies in specific cases. 

\begin{table}[h!]
\caption{\label{tab:nayf4_comparison}%
Comparison between the thermometric characterization of the same Ln$^{3+}$-host system (NaYF$_4$: 20\% Yb$^{3+}$ / 2\% Er$^{3+}$) for different works in the literature.
}

\begin{tabular}{ccccc}
\hline
\begin{tabular}[c]{@{}c@{}}Average \\ Particle\\ Size\end{tabular} & \begin{tabular}[c]{@{}c@{}}Temperature \\ Range (K)\end{tabular} & \begin{tabular}[c]{@{}c@{}} $\Delta E_\text{eff}$ \\ (cm$^{-1}$) \end{tabular} & $C_\text{eff}$ & Ref.    \\ \hline

\qty{2}{\micro\meter} - \qty{6}{\micro\meter}            & 160-300                                                          & 752                          & 8.06      & [\!\!\citenum{Zhou_2013}] \\
\qty{1.5}{\micro\meter}                                  & 295-325                                                          & 834                          & 14.88     & [\!\!\citenum{Goncalves_2021}] \\
\qty{1.1}{\micro\meter}                                  & 226-304                                                          & 379                          & 3.06      & [\!\!\citenum{Dong_2014}] \\
\qty{700}{\nano\meter}                                   & 226-304                                                          & 500                          & 4.89      & [\!\!\citenum{Dong_2014}] \\
\qty{70}{\nano\meter}                                    & 303-573                                                          & 748                          & 7.92      & [\!\!\citenum{Ding_2018}] \\
\qty{50}{\nano\meter}                                    & 300-400                                                          & 708                          & 14.12     & [\!\!\citenum{Pickel_2018}] \\
\qty{25}{\nano\meter}                                    & 296-400                                                          & 796                          & 5.8       & [\!\!\citenum{Kilbane_2016}] \\
\qty{20}{\nano\meter}                                    & 300-500                                                          & 730                          & 9.05      & [\!\!\citenum{Litao_2018}] \\ \hline

\label{tb:delta_E_comparison}
\end{tabular}

\end{table}

\section{Boltzmann luminescence thermometry with L\MakeLowercase{n}$^{3+}$ ions}

\subsection{Energy-level structure of $4f^N$-electron systems}

The Hamiltonian describing the $N$ electrons in a free $4f^N$ system encompasses the central potential interaction with the nucleus, interelectronic interactions among the $N$ electrons, spin-orbit coupling, and other terms of lower significance to the energy separation among states \cite{Malta_2003}. The nuclear central potential separates states in energy differences on the order of \qty{e5}{\per\centi\meter}, while the interelectronic repulsion and spin-orbit interaction contribute to separations on the order of \qty{e4}{\per\centi\meter} and \qty{e3}{\per\centi\meter}, respectively \cite{Malta_2003}. In Ln$^{3+}$ ions, the spin-orbit coupling is significant and cannot be treated as a small perturbation, making the intermediate coupling scheme (written on a basis of LS-coupling states) more suitable for describing the eigenstates as \cite{Malta_2003}

\begin{equation}
    \left|JM_J\right\rangle = \sum_{\gamma L S} C(\gamma L S) \left|(4f^N)\gamma[LS]JM_J\right\rangle \quad,
\label{eq:free_ion_eigenfunc}
\end{equation}

\noindent where $L$ and $S$ stand for the total orbital and spin angular momentum quantum numbers, respectively. $J$ stands for the total angular momentum quantum number, and $M_J$ is its projection in a given axis. $\gamma$ is a set of quantum numbers included when $L$, $S$, $J$ and $M_J$ do not unambiguously describe the eigenstates. The sum of $|C(\gamma L S)|^2$ over all indexes is equal to 1. For a given $J$, all $(2J+1)$ states for which $M_J = -J, -(J+1), ..., J-1, J$ are degenerate. Those degenerate subspaces are represented by the $^{2S+1}L_J$ manifold. For instance, the free Er$^{3+}$ ion has 11 electrons in the $4f$ shell. Its ground state manifold is $^{4}$I$_{15/2}$, which is 16-fold degenerate.

When considering the Ln$^{3+}$ ions embedded in crystals, the crystal field lifts the degeneracy of the total angular momentum quantum number, splitting the energy levels in the order of magnitude of \qty{e2}{\per\centi\meter} \cite{Malta_2003}. Not all degeneracy will be necessarily lifted. Thus, depending exactly on the crystal's symmetry arrangement, some crystal-field levels can still remain degenerate. Each crystal-field energy level will now be labeled by an irreducible representation of the symmetry group. In a first approximation, the eigenfunctions will be a sum over different $\left|JM_J\right\rangle$ for the same $J$. In a second-order correction, the ligand field may mix states with different $J$ (J-mixing effect), although this is a very small effect for most of Ln$^{3+}$ ions \cite{Malta_2003}. Furthermore, for ions with half-integer $J$ in all crystal environments except cubic, the crystal-field levels are Kramer's doublets, meaning that each level is doubly degenerate \cite{Hanninen_2010}. As a result, the ground state of the Er$^{3+}$ ion in a tetragonal crystal site, for instance, will have 8 crystal-field levels. These levels are also called Stark sublevels because the host environment can be viewed as producing a static electric field on the ion's position, to first order.

\subsection{Boltzmann distribution: A two-level system with phonon-mediated interactions}

Atoms in a crystal oscillate around an equilibrium position, modulating the electric field at the Ln$^{3+}$ ion's site and enabling electron-phonon interactions. As a result, electronic population can undergo nonradiative transitions between two vibronic levels through phonon-mediated interactions. Such processes can be described as a $p$-order approximation process, where $p$ phonons of an effective phonon mode are created or annihilated \cite{Layne_1977, Riseberg_1968}. This effective phonon mode has energy below the phonon cutoff energy and bridges the energy gap between the vibronic levels with the smallest $p$ \cite{Menezes_2001}.

Consider a two-level system embedded in a crystal field. The quantum states are $|a\rangle$ and $|b\rangle$, with energies $E_b > E_a$. The non-radiative (phononic) emission and absorption rates between those levels can be expressed as $g_a W_{\text{NR}}^{(0)} \left(1+ \langle n_\text{eff}\rangle\right)^p$ and $g_b W_{\text{NR}}^{(0)} \langle n_\text{eff}\rangle^p$, respectively \cite{Suta_Meijerink_2020}, where $g_a$ and $g_b$ are the degeneracy of the levels $|a\rangle$ and $|b\rangle$, respectively. $W_{\text{NR}}^{(0)}$ is the spontaneous phonon emission rate (at zero temperature) and $\langle n_\text{eff}\rangle$ is the thermal average occupation number of the effective phonon mode, which presents a temperature dependence given by Bose-Einstein statistics 

\begin{equation}
\langle n_\text{eff}\rangle = \left[\exp{\left(\beta \,\hbar \omega_\text{eff}\right)}-1\right]^ {-1} \; ,
	\label{eq:non_rad_decay}
\end{equation}

\noindent being $\beta = (k_\mathrm{B} T)^{-1}$ and $\omega_\text{eff}$ the phonon mode angular frequency. The explicit dependence of the non-radiative decay and absorption rates with the temperature are, respectively \cite{Menezes_2003}:

\begin{equation}
	W_{b \rightarrow a}^{\text{dec}} (T) = g_a W_{\text{NR}}^{(0)}\left[ 1-\exp{\left(\frac{-\hbar \omega_\text{eff}}{k_\mathrm{B} T}\right)} \right]^{-p} \;,
	\label{eq:non_rad_decay}
\end{equation}

\begin{equation}
	W_{a \rightarrow b}^{\text{abs}} (T) = g_b W_{\text{NR}}^{(0)}\left[\exp{\left(\frac{\hbar \omega_\text{eff}}{k_\mathrm{B} T}\right)}-1 \right]^{-p}\; .
	\label{eq:non_rad_absorp}
\end{equation}

If we now consider an ensemble of $N^\prime$ non-interacting two-level systems, where the only electronic exchange between $|a\rangle$ and $|b\rangle$ is through a $p$-order phonon-mediated process, for which the emission and absorption transition rates are given by Eqs. \eqref{eq:non_rad_decay} and \eqref{eq:non_rad_absorp}, respectively, the rate equation for their populations ($n_i$) can be written as

\begin{equation}
\begin{aligned} 
	\dot{n_b} &= n_a \cdot W_{a \rightarrow b}^{\text{abs}} (T) - n_b \cdot W_{b \rightarrow a}^{\text{dec}} (T) \\
    n_a &= N^\prime - n_b \; .
\end{aligned} 
	\label{eq:phonon_rate_equation} 
\end{equation}

\noindent By solving it in steady-state and substituting Eqs. \eqref{eq:non_rad_decay} and \eqref{eq:non_rad_absorp}, one finds the population ratio satisfies

\begin{equation}
	\frac{n_b}{n_a} = \frac{W_{a \rightarrow b}^{\text{abs}} (T)}{W_{b \rightarrow a}^{\text{dec}} (T)} = \frac{g_b}{g_a} \exp{\left(-\frac{E_b - E_a}{k_{\text{B}}T}\right)},
	\label{eq:phonon_LIR}
\end{equation}

\noindent where $E_b - E_a = p\,\hbar \omega_\text{eff}$ implies the restriction that the energy of the $p$ effective phonons must match the energy difference between the levels. From the macroscopic point of view, Eq. \eqref{eq:phonon_LIR} represents exactly the Boltzmann distribution, retrieved when considering that the ions constitute a canonical ensemble in contact with a thermal reservoir of temperature $T$ \cite{Suta_Meijerink_2020}.

\subsection{Boltzmann luminescence thermometry using two crystal-field levels}

In Ln$^{3+}$ ions, the two-level system interacting solely through the matrix's phonons can be experimentally approximated when the phonon-mediated transition rates between them dominate over all other photophysical processes. In this case, there is a rapid thermalization between the states, generally in the order of a few microseconds to nanoseconds \cite{Dechao_2016}, retrieving the Boltzmann population distribution. The energy levels satisfying such requirements are TCLs and the energy difference between them must be comparable to the matrix average phonons' energies \cite{Suta_Meijerink_2020}. By having access to the population ratio of the TCLs, one could retrieve the absolute temperature from Eq. \eqref{eq:phonon_LIR}. In Ln$^{3+}$ ions, the `TCLs' nomenclature is used even for manifolds under the above-mentioned conditions, therefore we will use the term TCLs also for thermally coupled manifolds throughout this manuscript. For manifolds, the thermalization will take place among all their Stark sublevels.

\begin{figure} 
	\includegraphics[width=\linewidth]{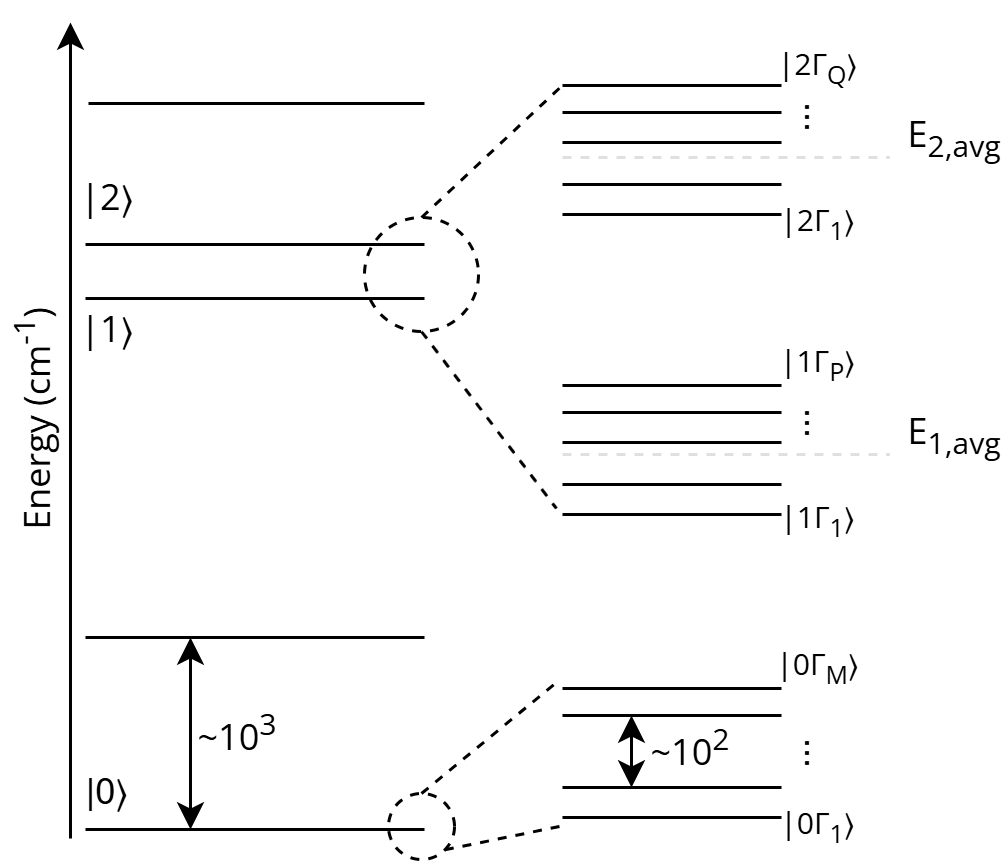}
	\caption{Energy-level diagram of Ln$^{3+}$ ions doped in crystalline hosts. The spin-orbit manifolds are represented by $|i\rangle$ and its Stark sublevels by $|i\Gamma_{k}\rangle$.}
	\label{fig:energy_levels}
\end{figure}

Figure \ref{fig:energy_levels} shows a typical energy level diagram for Ln$^{3+}$ ions. The $|0\rangle$, $|1\rangle$ and $|2\rangle$ are spin-orbit manifolds. We consider that $|1\rangle$ and $|2\rangle$ are TCLs. For each $|i\rangle$ manifold, the crystal field splits it into $|i\Gamma_{k}\rangle$ sublevels ($k = \{1,..,K\}$) with energy $E_{ik}$. The average energy of the $|i\rangle$ manifold is 

\begin{equation}
    E_{i,{\text{avg}}} = \frac{\sum_{k=1}^{K} E_{i,k}}{K} \; ,
    \label{eq:Eiavg_def}
\end{equation}

\noindent which runs over the $K$ crystal-field levels. In the approximation that the TCLs' population (including all their sublevels) follow Boltzmann population distribution, one finds that the probability of a given level $|i\Gamma_{k}\rangle$ being occupied is 

\begin{equation}
	p_{ik}(T) = \frac{g_{ik} \exp{\left(-\frac{E_{ik}-E_{11}}{k_{\text{B}}T}\right)}}{\mathlarger{\sum}_{j=\{1,2\}}\mathlarger{\sum}_{l=1}^{L_j} g_{jl}\exp{\left(-\frac{E_{jl}-E_{11}}{k_{\text{B}}T}\right)}} \; ,
	\label{eq:boltzmann_distribution}
\end{equation}

\noindent where $g_{ik}$ is the degeneracy of $|i\Gamma_{k}\rangle$. Notice that the sum in the denominator runs over all crystal-field levels of both thermally coupled manifolds. We have considered the `ground' state for the thermalization  as the lowest Stark sublevel, $|1\Gamma_{1}\rangle$.

Experimentally, the TCLs are commonly populated through light absorption, generally using upconversion or downconversion excitation scheme in a way that the excitation laser does not spectrally overlap with luminescence bands used for performing the thermometry measurements, thus avoiding experimental artifacts \cite{Goncalves_2021, Galvao2_2021}. Once the TCLs are populated, the excited ions can radiatively decay to lower-lying states, emitting photons. The spontaneous radiative decay rate must be much smaller than the phonon-induced thermalization to sustain Boltzmann distribution. 

Consider the spontaneous radiative transition $|i\Gamma_k\rangle \rightarrow |j\Gamma_l\rangle$, where $|i\Gamma_k\rangle$ is a Stark sublevel of the TCLs and $|j\Gamma_l\rangle$ is a Stark sublevel of a lower-lying manifold, generally the ground state. The number of photons per unit time emitted in this transition is 

\begin{equation}
\Phi_{ik,jl} = n_{ik} \,A_{ik,jl} \quad ,
	\label{eq:photons_emitted}
\end{equation}

\noindent where $n_{ik}$ is the number of excited ions in the level $|i\Gamma_k\rangle$, given by $N^\prime p_{ik}(T)$ (Eq. \ref{eq:boltzmann_distribution}), with $N^\prime$ being the total electronic population in the TCLs (amount of excited ions in the system). $A_{ik,jl}$ is the Einstein coefficient for the spontaneous Stark-Stark emission. In the dipole approximation, it is given by \cite{Axner_2004}

\begin{equation}
	A_{ik,jl} = \frac{4\alpha\omega^3_{ik,jl}}{3c^2 \, g_{ik}} \left[ n |\chi(n)|^2 S_{ED} + n^3 S_{MD}\right],
	\label{eq:spontaneous_rate}
\end{equation}

\noindent where $\alpha$ is the fine structure constant, $\omega_{ik,jl}$ is the frequency of the photon emitted in the transition, $n$ is the index of refraction of the host medium and $|\chi(n)|^2$ is a local electric field correction factor \cite{Suta_Meijerink_2020}. $S_{ED}$ and $S_{MD}$ are the electric and magnetic dipole line strengths, respectively, given by \cite{Leavitt_1980}

\begin{equation}
	S_{ED} = \sum_{a,b} |\langle j\Gamma_l,b| \, \sum_n \mathbf{r}_n \, |i\Gamma_k,a\rangle |^2 \, ,
	\label{eq:ED_line_strength}
\end{equation}

\begin{equation}
	S_{MD} = \left(\frac{\mu_\text{B}}{e\,c}\right)^2
\sum_{a,b} \left|\langle j\Gamma_l,b| \, \mathbf{L} + g_\text{S}\mathbf{S} \,|i\Gamma_k,a\rangle\right|^2 \, .
	\label{eq:MD_line_strength}
\end{equation}

\noindent $|i\Gamma_k,a\rangle$ represents each individual quantum state which composes the level $|i\Gamma_k\rangle$. Similarly to $|j\Gamma_l,b\rangle$. The sum over $n$ in $S_{ED}$ runs for all $4f^N$ electrons of the Ln$^{3+}$ ions, being $\mathbf{r}_n$ their respective positions. In $S_{MD}$, $\mu_\text{B}$ is Bohr's magneton, $e$ the electron charge, $\mathbf{L}$ is the sum over all $4f^N$ electron' orbital angular momentum, and $\mathbf{S}$ the spin counterpart. $g_\text{S} = 2 + \mathcal{O}(10^{-3})$ is the g-factor for electrons. Notice that according to the definition of Eqs. \eqref{eq:spontaneous_rate}, \eqref{eq:ED_line_strength} and \eqref{eq:MD_line_strength}, the total transition rate between manifolds $|i\rangle \rightarrow |j\rangle$ can be defined by the sum \cite{Leavitt_1980}

\begin{equation}
	g_{i} \, A_{ij} = \mathlarger{\sum}_{k,l} A_{ik,jl}\, g_{ik}
 \label{eq:sum_line_to_line}
\end{equation}

\noindent where $g_{i} = (2J_i + 1)$ is the degeneracy of the $|i\rangle$ multiplet. The manifold-to-manifold transition rates $A_{ij}$ can be obtained from the Judd-Ofelt theory, assuming that all Stark sublevels of the given manifold are equally populated \cite{Walsh_2006}.

Now consider that the light emission from the radiative transition $|i\Gamma_k\rangle \rightarrow |j\Gamma_l\rangle$ has a lineshape function $L_{ik,jl}(\lambda)$, such that $\int L_{ik,jl}(\lambda) \, d\lambda = 1$. Thus, the wavelength distribution of the photon emission rate can be written as $\Phi_{ik,jl}L_{ik,jl}(\lambda)$. Also consider that one is using a spectrometer with photocounting detectors to measure the luminescence spectrum. A fraction of the emitted photons will impinge on the photodetector's cross-section unit ($d\lambda$), which can be detected with a certain efficiency $\eta(\lambda)$, that must include the efficiencies of the optical components along the optical path, including the photodetector itself. The photon-count rate measured by each detector's cross-section unit will be \cite{Suta_Meijerink_2020} $\eta(\lambda)\Phi_{ik,jl}L_{ik,jl}(\lambda) \, d\lambda$. If the line-shape function $L_{ik,jl}(\lambda)$ falls off rapidly enough from the central wavelength, the integration interval can be limited to $[\lambda_1, \lambda_2]$. Also, the spectrometer resolution typically allows to assume that $\eta(\lambda)$ does not change significantly in the wavelength interval considered, thus $\eta(\lambda) \approx \eta(\tilde\lambda_{12})$, being $\tilde\lambda_{12}$ the average wavelength of the integration range. By integrating the detected spectrum over the detector's cross-section units in the interval $[\lambda_1, \lambda_2]$, the transition photon-count rate results in $\eta(\tilde\lambda_{12})\Phi_{ik,jl}$.

To retrieve the temperature information from the luminescence spectrum, one can observe the radiative decay from two different Stark sublevels of the TCLs to a lower-lying Stark state (\textit{e.g.} the ground state $|0\Gamma_m\rangle$). Consider that the observed transitions are $|i\Gamma_k\rangle \rightarrow |0\Gamma_m\rangle$ and $|j\Gamma_l\rangle \rightarrow |0\Gamma_{m'}\rangle$, with energies $E_{ik}>E_{jl}$. If those spectral lines do not spectrally overlap, one can set different integration intervals, say $[\lambda_1, \lambda_2]$ for the upper thermally coupled band, and $[\lambda_3, \lambda_4]$ for the lower thermally coupled band. The LIR ($R$) will be given by 

\begin{equation}
\begin{aligned}
R(T)&=\frac{\eta(\tilde\lambda_{12})\mathlarger{\int}_{\lambda_1}^{\lambda_2}\Phi_{ik,0m} \, L_{ik,0m}(\lambda)\,d\lambda}{\eta(\tilde\lambda_{34}) \mathlarger{\int}_{\lambda_3}^{\lambda_4} \Phi_{jl,0m'}\, L_{jl,0m'}(\lambda) \, d\lambda} = \frac{A_{ik,0m} n_{ik}}{A_{jl,0m'} n_{jl}} \\
 &= \frac{A_{ik,0m}}{A_{jl,0m'}} \frac{g_{ik}}{g_{jl}} \cdot \exp{\left(-\frac{\Delta E_{ik,jl}}{k_{\text{B}}T}\right)} \; ,
 \end{aligned}
 	\label{eq:LIR_Stark_Stark}
\end{equation}

\noindent where we have considered the situation in which the detection quantum efficiency does not change in the region of interest ($\eta(\tilde\lambda_{12}) \approx \eta(\tilde\lambda_{34})$). If this is not the case, the efficiencies can be embedded in the decay rate constants, defining effective values. This general formula is valid even for sublevels inside the same manifold ($i=j$), which can be interesting for low-temperature applications, where $k_\text{B}T$ is of the order of the energy-level separation.

From Eq. \eqref{eq:LIR_Stark_Stark}, it can be concluded that if one can identify and spectrally separate the emissions from single Stark-Stark transitions, the microscopic parameters, ${A_{ik,0m}}/{A_{jl,0m'}}$ and $\Delta E_{ik,jl}$, are directly accessible through a $R(T)$ \textit{vs}. $T$ experimental curve. Furthermore, by calculating \textit{ab-initio} $A_{ik,0m}$ and $A_{jl,0m}$ (or somehow measuring it), and knowing $\Delta E_{ik,jl}$ with some accuracy, it should be possible to build a truly primary thermometer, independent of calibration by an external thermometer.

\subsection{Boltzmann luminescence thermometry using spin-orbit manifolds}

Despite the simplicity of Eq. \eqref{eq:LIR_Stark_Stark} and its direct connection to the microscopic parameters, the challenge with performing ratiometric thermometry with crystal-field levels remains because it is not straightforward to separate all Stark-Stark lines in the spectrum, especially at room temperature. A much more common approach in the literature of Ln$^{3+}$-based ratiometric thermometry is to consider the complete manifold-to-manifold transition in the wavelength (or energy) integration to calculate $R(T)$. However, as we will show in the following, the LIR's temperature dependence is different from that showed in Eq. \eqref{eq:LIR_Stark_Stark}.

Consider that the two manifold-to-manifold spectral bands $|2\rangle \rightarrow |0\rangle$ and $|1\rangle \rightarrow |0\rangle$ do not overlap in wavelength, thus one can integrate the spectrum over $[\lambda_1^\prime, \lambda_2^\prime]$ for the former and $[\lambda_3^\prime, \lambda_4^\prime]$ for the latter. The $|i\rangle \rightarrow |0\rangle$ band contains all Stark-Stark transitions. For example, in Er$^{3+}$ ions, the transition $^{2}$H$_{11/2}$ $\rightarrow$  $^{4}$I$_{15/2}$ could possibly have $6\times 8 = 48$ Stark-Stark lines, depending of course on the selection rules, which are embedded in the $A_{ik,0m}$ factors. The LIR obtained by integrating each manifold in this case is

\begin{equation}
\begin{aligned}
R(T)&=\frac{\mathlarger{\int}_{\lambda_1^\prime}^{\lambda_2^\prime}\mathlarger{\sum}_{km}\, \Phi_{2k,0m} \, L_{2k,0m}(\lambda)\, d\lambda}{\mathlarger{\int}_{\lambda_3^\prime}^{\lambda_4^\prime} \mathlarger{\sum}_{lm}\, \Phi_{1l,0m} \, L_{1l,0m}(\lambda) \, d\lambda} = \frac{\mathlarger{\sum}_{km}\, A_{2k,0m} \, n_{2k}}{\mathlarger{\sum}_{lm}\, A_{1l,0m} \, n_{1l}} \\
 &= \frac{\mathlarger{\sum}_{km}\, A_{2k,0m} \, g_{2k} \, \exp{\left(-E_{2k}/k_{\text{B}}T\right)}}{\mathlarger{\sum}_{l m}\, A_{1l,0m} \, g_{1l} \, \exp{\left(-E_{1l}/k_{\text{B}}T\right)}} \; ,
\end{aligned}
\label{eq:LIR_maifold_manifold}
\end{equation}

\noindent where $k$, $l$ and $m$ run over all crystal-field levels of $|2\rangle$, $|1\rangle$ and $|0\rangle$, respectively. Since each Stark level has a different occupation probability and each one has its own oscillator strengths for transitions to another Stark level of the ground state, they contribute differently to the temperature dependence of the LIR. 

Writing the LIR's temperature dependence as Eq. \eqref{eq:LIR_maifold_manifold} requires knowing all Stark-Stark oscillator strengths between the two manifolds to the ground state, which is a challenging task. Instead, the standard approach in the Ln$^{3+}$-based Boltzmann thermometry community is to calibrate the system before its usage \cite{Goncalves_2021, Galvao2_2021}. The procedure consists in obtaining a set of luminescence spectra at different temperatures $T$, measured by an external thermometer, and then one fits the curve $\ln(\text{LIR})$ \textit{vs.} $T^{-1}$ by the linearized relation

\begin{equation}
	\ln R(T) = \ln C_\text{eff} -\frac{\Delta E_\text{eff}}{k_{\text{B}}T} \quad,
	\label{eq:LIR_fitting}
\end{equation}

\noindent where $C_\text{eff}$ and $\Delta E_\text{eff}$ are fitting parameters, supposedly constant in the temperature range, also called calibration parameters. In the situation where the Stark splittings are very small compared to $k_{\text{B}}T$, then $R(T)$ in Eq. \eqref{eq:LIR_maifold_manifold} can be well-approximated by a single-exponential function with the prefactor as $g_2A_{20}/g_1A_{10}$ (by using Eq. \ref{eq:sum_line_to_line}) and one restores the temperature dependence of Eq. \eqref{eq:LIR_fitting}. However, it is worth noticing that $k_{\text{B}}T$ at room temperature is in the same order of magnitude as the Stark separation between the levels, \qty{e2}{\per\centi\meter}\cite{Kisliuk_1964}. As a result, it will be shown below that higher-order corrections become increasingly important. Another important consequence is that the calibration parameters do not have a straightforward relation to the microscopic quantities, \textit{e.g.} $A_{ik,0m}$ and $\Delta E_{ik,0m}$, even though the fitting R-squared factor is close to 1 for a sufficiently low-temperature range. It was previously shown by Galindo \textit{et al.} \cite{Galindo_2021} that the experimentally retrieved parameters $C_\text{eff}$ and $\Delta E_\text{eff}$ can be strongly correlated, meaning that any microscopic parameter can be challenging to determine from LIR measurements directly.

\subsubsection{Microscopic origin of $C_\text{eff}$ and $\Delta E_\text{eff}$}

To understand the microscopic origin of the calibration parameters, one can consider that the thermometry experiments are performed under a sufficiently small temperature range and expand Eq. \eqref{eq:LIR_maifold_manifold} in terms of $\beta = (k_\text{B}T)^{-1}$ up to first order around the central $\beta_c = (k_\text{B}T_c)^{-1}$, where $T_c$ is the central temperature of the range. Hence, one gets

\begin{equation}
	\ln R(T) \approx \left[\ln R(T_c) + T_c\cdot S_r(T_c)\right]-\frac{S_r(T_c) \cdot k_\text{B}T_c^2}{k_{\text{B}}T} \quad,
	\label{eq:lnLIR_expansion_Taylor}
\end{equation}

\noindent where $S_r$ is the thermometer's relative sensitivity, defined as

\begin{equation}
S_r = \frac{1}{R}\frac{\partial R}{\partial T} \; .
\label{eq:def_sens_rel}
\end{equation}

\noindent Therefore, the experimentally-measured calibration parameters are 

\begin{equation}
\begin{aligned}
&\ln C_\text{eff} = \ln R(T_c) + T_c\cdot S_r(T_c) \\
&\Delta E_\text{eff} = S_r(T_c) \cdot k_\text{B}T_c^2 \quad , 
\label{eq:DeltaE_eff_andC_eff}
\end{aligned}
\end{equation}

\noindent where $S_r$ is calculated based on Eq. \eqref{eq:LIR_maifold_manifold}. This is an important result showing the connection between $\ln C_\text{eff}$ and $\Delta E_\text{eff}$ with the microscopic quantities $A_{ik,0m}$ and $\Delta E_{ik,0m}$. Also, it shows how the measured values change depending on the chosen experimental temperature range. Furthermore, since both $\ln C_\text{eff}$ and $\Delta E_\text{eff}$ depend on $S_r$, it may lead to correlation between those variables, especially when the temperature range is small. Such correlation may be reduced for experiments across a high temperature range, where the finer details of Eq. \eqref{eq:LIR_maifold_manifold} may become more prominent and helpful to distinguish the effective parameters.

By taking the natural logarithm of Eq. \eqref{eq:LIR_maifold_manifold}, we observe that $\ln R(T)$ can be expressed in terms of weighted cumulant-generating functions \cite{Kenney_1951} $K_{t}(\mathbf{x},\mathbf{y}) = \ln \sum_k y_k \exp{(t\, x_k)}$, which is also called weighted LogSumExp function in the machine learning and neural network community \cite{Schiele_2024}. Eq. \eqref{eq:LIR_maifold_manifold} becomes

\begin{equation}
\begin{aligned}
\ln{R(T)} = \ln{\left(\frac{g_2 \, A_{20}}{g_1 \, A_{10}} \right)} &+ K_{-\beta}(\mathbf{E}_2, \boldsymbol{\mathcal{\mathit{w}}}_{20}) \\ & - K_{-\beta}(\mathbf{E}_1, \boldsymbol{\mathcal{\mathit{w}}}_{10}) \, .
\end{aligned}
\label{eq:LIR_Cumulant_generating}
\end{equation}

\noindent The vector $\boldsymbol{\mathcal{\mathit{w}}}_{i0}$ is normalized, having its $k$-th element given by

\begin{equation}
\mathcal{\mathit{w}}_{i0,k} = \mathcal{\mathit{w}}_{ik} = \frac{\sum_m g_{ik}\, A_{ik,0m}}{g_iA_{i0}} \; ,
\label{eq:def_wik}
\end{equation}

\noindent where we have made the zero subscript implicit. Similarly, the $k$-th element of $\mathbf{E}_i$ is $E_{ik}$. Notice that $\sum_k \mathcal{\mathit{w}}_{ik} = 1$. The identification of the cumulant-generating functions is important because the coefficients of the power-series expansion of $K_{-\beta}(\mathbf{E}_i, \boldsymbol{\mathcal{\mathit{w}}}_{i0})$ around infinite temperature ($\beta \rightarrow 0$) give the cumulants of the set $\{E_{ik}\}$ weighted by $\boldsymbol{\mathcal{\mathit{w}}}_{i0}$ \cite{Kenney_1951}. The cumulants connect directly to the central moments of the distribution (average, variance, skewness, etc), which in this case will be all weighted by the levels' degeneracy and oscillator strengths. Up to first-order in $\beta$, one gets

\begin{equation}
\begin{aligned}
\ln{R(T \rightarrow \infty)} &= \ln{\left(\frac{g_2 \, A_{20}}{g_1 \, A_{10}} \right)} - \frac{(\overline{\mathrm{E}_2}_\mathcal{\mathit{w}}-\overline{\mathrm{E}_1}_\mathcal{\mathit{w}})}{k_{\text{B}}T} \; ,
\end{aligned}
\label{eq:ln_LIR_cumulant}
\end{equation}

\noindent where $\overline{\mathrm{E}_i}_\mathcal{\mathit{w}} = \sum_{k}\, \mathcal{\mathit{w}}_{ik} \, E_{ik}$ is the weighted average of the Stark  sublevels' energy of the $|i\rangle$ manifold. The next term of the expansion would include the second central moment $\sigma_{i\mathcal{\mathit{w}}}^2 = \sum_{k}\, \mathcal{\mathit{w}}_{ik} \, \, E_{ik}^2 - \overline{\mathrm{E}_i}_\mathcal{\mathit{w}}$, which is related to the bandwidth of the luminescence spectrum. Notice that $\overline{\mathrm{E}_2}_\mathcal{\mathit{w}}-\overline{\mathrm{E}_1}_\mathcal{\mathit{w}}$ corresponds to the infinite-temperature limit for $\Delta E_\text{eff}$, as we will show in the following section.

\subsubsection{Relation between $\Delta E_\text{eff}$ and $\Delta E_{\text{bary}}$}

We are now able to investigate the interconnection between $\Delta E_\text{eff}$ and other related quantities, such as i) the difference between the average energies of the Stark sublevels, $\Delta E_{\text{avg}}$, calculated based on the energy-level structure (Eq. \ref{eq:Eiavg_def}); and ii) the difference between the energy barycenters of the TCLs' luminescence bands, $\Delta E_{\text{bary}}$, calculated based on the luminescence spectrum. 

Firstly, by using Eq. \eqref{eq:def_sens_rel} in Eq. \eqref{eq:DeltaE_eff_andC_eff}, developing the derivatives with respect to $T$, and using the weights defined in Eq. \eqref{eq:def_wik}, one gets

\begin{equation}
\begin{aligned}
\Delta E_\text{eff} = \frac{ \sum_k E_{2k}\mathcal{\mathit{w}}_{2k} \; e^{-\beta_c\, E_{2k}}}{\sum_k \mathcal{\mathit{w}}_{2k} \, e^{-\beta_c\, E_{2k}}} 
 - \frac{\sum_l E_{1l}\mathcal{\mathit{w}}_{1l} \; e^{-\beta_c\, E_{1l}}}{\sum_l \mathcal{\mathit{w}}_{1l} \, e^{-\beta_c\, E_{1l}}} \; .
\label{eq:DeltaE_eff}
\end{aligned}
\end{equation}

\noindent On the other hand, the barycenter of the luminescence band resulting from the transition $|i\rangle \rightarrow |0\rangle$ averages the energies of all Stark-Stark lines by their strengths. It should be calculated in the energy space \cite{Mooney_2013}. The lineshape function in the energy space,  $\tilde L_{ik,0m}(E)$, is related to $L_{ik,0m}(\lambda)$ by \cite{Suta_Meijerink_2020} 

\begin{equation}
\tilde L_{ik,0m}(E) = \left(\frac{\lambda^2}{hc}\right)L_{ik,0m}(\lambda) \,.
\label{eq:energy_wvl_transf}
\end{equation}

\noindent Therefore, the barycenter of $|i\rangle \rightarrow |0\rangle$ can be written as

\begin{equation}
\begin{aligned}
&E_{i,\text{bary}} = \frac{\mathlarger{\int}_{E_1^\prime}^{E_2^\prime}E \; \mathlarger{\sum}_{km}\, \Phi_{ik,0m} \, \tilde L_{ik,0m}(E)\, dE}{\mathlarger{\int}_{E_1^\prime}^{E_2^\prime}\mathlarger{\sum}_{km}\, \Phi_{ik,0m} \, \tilde L_{ik,0m}(E)\, dE} \\
&=\frac{\mathlarger{\sum}_{km}\, \left( E_{ik} - E_{0m} \right) A_{ik,0m} \, g_{ik}\exp{\left(-E_{ik}/k_\text{B} T \right)}}{\mathlarger{\sum}_{km}\, A_{ik,0m}g_{ik}\exp{\left(-E_{ik}/k_\text{B} T \right)}} \; .
\end{aligned}
\label{eq:barycenter_manifold_energy}
\end{equation}

\noindent The quantity $E_{ik,0m} = E_{ik} - E_{0m}$ is the result of the integration $\int E \tilde L_{ik,0m}(E) \, dE$, which is the energy of the photon emitted in the transition $|i\Gamma_k\rangle \rightarrow |0\Gamma_m\rangle$. Notice that since the barycenter depends on $\Phi_{ik,0m}$ it also depends on the Stark levels' populations, which vary with the temperature. As the temperature increases, higher-lying Stark sublevels of a given manifold become more populated. They will contribute with more high-energy photons to the spectrum. Consequently, $E_{i,\text{bary}}$ tends to increase with increasing temperature.

By using Eq. \eqref{eq:barycenter_manifold_energy}, the quantity $\Delta E_\text{bary}$ can be written as $E_{2,\text{bary}} - E_{1,\text{bary}}$. We notice that, at $T=T_c$, the relation between $\Delta E_\text{eff}$ and $\Delta E_{\text{bary}}$ is

\begin{strip}
\begin{center}
\begin{equation}
\begin{aligned}
\Delta E_\text{bary} = \Delta E_{\text{eff}}
- \left(\frac{\sum_{km}\, E_{0m} A_{2k,0m} \, g_{2k}e^{-\beta_c\, E_{2k}}}{\sum_{km}\, A_{2k,0m}g_{2k}e^{-\beta_c\, E_{2k}}} \right.
\left.- \frac{\sum_{lm}\, E_{0m} A_{1l,0m} \, g_{1l}e^{-\beta_c\, E_{1l}}}{\sum_{lm}\, A_{1l,0m}g_{1l}e^{-\beta_c\, E_{1l}}}\right)\; .
\end{aligned}
\label{eq:DeltaEeff_vs_DeltaEbary}
\end{equation}
\end{center}
\end{strip}

\noindent Showing that $\Delta E_\text{eff}$ and $\Delta E_{\text{bary}}$ are conceptually different quantities. The quantitative difference depends exactly on the Stark energy level structure and on the Stark-Stark line strengths between the TCLs and the ground state. Nevertheless, if one considers only transitions from the TCLs to a single Stark sublevel of the ground state in the integration of Eq. \eqref{eq:LIR_maifold_manifold} (or, equivalently, if the ground state is completely degenerate), then $\Delta E_\text{eff}$ and $\Delta E_{\text{bary}}$ will be equal. 

At infinite temperature, $\Delta E_\text{eff}$ becomes $\sum_k (\mathit{w}_{2k}E_{2k} - \mathit{w}_{1k}E_{1k})$, which is equal to $\overline{\mathrm{E}_2}_\mathcal{\mathit{w}}-\overline{\mathrm{E}_1}_\mathcal{\mathit{w}}$ as in Eq. \eqref{eq:ln_LIR_cumulant}. Conversely, $\Delta E_{\text{bary}}$ becomes

\begin{equation}
\begin{aligned}
&\Delta E_\text{bary} (T\rightarrow\infty) = \Delta E_\text{eff} (T\rightarrow\infty) 
\\ &- \left(\frac{\sum_{km}  E_{0m} A_{2k,0m} \, g_{2k}}{g_2 A_{20}} - \frac{\sum_{lm}  E_{0m} A_{1l,0m} \, g_{1l}}{g_1 A_{10}} \right) \; ,
\end{aligned}
\label{eq:deltaE_21_bary_atTinf}
\end{equation}

\noindent demonstrating that $\Delta E_\text{eff}$ and $\Delta E_{\text{bary}}$ are different regardless of the temperature for a general set of $A_{ik,0m}$. The general temperature dependence of the energy barycenter is given by Eq. \eqref{eq:barycenter_manifold_energy}, but if we consider a simplified case where $A_{ik,0m}$ are equal for all Stark-Stark transitions (despite not physically likely, this model is useful to draw physical insights), then $\Delta E_\text{eff} = \Delta E_{\text{bary}}$ for all temperatures. Also, for $T \rightarrow \infty$, $\Delta E_{\text{bary}}$ simply becomes the average energy difference between the TCLs, $\Delta E_{\text{avg}}$ (Eq. \ref{eq:Eiavg_def}). For a general set of $A_{ik,0m}$, one cannot predict whether $\Delta E_{\text{bary}}$ is greater or lower than $\Delta E_{\text{avg}}$ since the line strengths can push the barycenter to higher or lower energy values. Therefore, an absolute calibration of luminescent thermometers based on transitions involving complete spin-orbit-manifolds is difficult in general because it is essential to know the oscillator strengths for each radiative transition between the manifolds' Stark sublevels. 

\subsubsection{Primary ratiometric thermometry}

In 2017, Balabhadra \textit{et al.} developed a method \cite{Balabhadra_2017} based on Eq. \eqref{eq:LIR_fitting} in which the $C_\text{eff}$ value is obtained based on a LIR \textit{vs.} Excitation Power curve and $\Delta E_\text{eff}$ is considered as equal to $\Delta E_\text{bary}$ at room temperature. For a known room temperature (measured with an external thermometer), the authors measured the LIR for various laser excitation powers and linearly extrapolated the observed dependence for zero power excitation, making it possible to estimate the $C_\text{eff}$ parameter. This calibration is necessary to minimize the effects of self-heating due to the laser power, which can lead to different responses on different surrounding media. Also, since it is based on an external temperature measurement, this calibration process embeds the correlation between the measured $C_\text{eff}$ and $\Delta E_\text{bary}$, allowing its usage despite its difference to $\Delta E_\text{eff}$, as shown in the previous section. Despite being an excellent method for accounting for thermometers in different media, the method still needs an external temperature measurement as a prior calibration step.

Alternatively, other authors employed the Judd-Ofelt theory to obtain $C_\text{eff}$ \cite{LeonLuis_2012, Rakov_2017}. As already mentioned, when one performs the sum of Eq. \eqref{eq:sum_line_to_line}, the manifold-to-manifold transitions rates, $A_{ij}$, can be written in the Judd-Ofelt theory framework as \cite{Walsh_2006}

\begin{equation}
A_{ij} = \mathlarger{\sum}_{\lambda = 2,4,6}\Omega_\lambda
\left\langle{ }j\left\|U^{(\lambda)}\right\|i\right\rangle^2 \;,
\label{eq:judd_ofelt_def}
\end{equation}

\noindent where $\Omega_\lambda$ are the Judd-Ofelt parameters, generally determined experimentally from light absorption measurements. The reduced matrix elements $\left\langle{ }j\left\|U^{(\lambda)}\right\|i\right\rangle$ are tabulated for each manifold-to-manifold transition for each Ln$^{3+}$ ion \cite{Carnall_1978}. Nexha \textit{et al.} compared the experimentally-obtained temperature calibration parameters with those obtained `theoretically' \cite{Nexha_2022}. In the former approach, $C_\text{eff}$ and $\Delta E_\text{eff}$ are determined through the LIR \textit{vs.} T curve fitting, while in the latter, $C_\text{eff}$ is calculated by using Eq. \eqref{eq:judd_ofelt_def} and $\Delta E_\text{eff}$ as the energy barycenter at room temperature. The authors have found that by using the `theoretical' parameters ($C_\text{eff}$ by Judd-Ofelt and $\Delta E_\text{eff}$ by the energy barycenter), the measured temperature is approximately \qty{40}{\kelvin} higher than that measured with the experimental parameters, for the same LIR value. They have attributed such temperature rise to the laser-induced heating, which is embedded in the experimental calibration for a given laser excitation power. However, as shown by Pickel \textit{et al.}, the upconversion process can also lead to non-Boltzmann distortion of the TCLs' population, therefore increasing the LIR, not necessarily connected to an increase of the particle's temperature \cite{Pickel_2018}.

In addition to possible laser-induced heating, or any dynamical effect, we must consider, more fundamentally, the fact that the correct temperature dependence is Eq. \eqref{eq:LIR_maifold_manifold} and not Eq. \eqref{eq:LIR_fitting}. And when using Eq. \eqref{eq:LIR_fitting}, the fitting parameters do not have a straightforward connection to the $A_{i0}$ and $\Delta E_{\text{bary}}$. In the next section, we provide simulations to quantify the order of magnitude of such differences. Another issue to consider is that, as already mentioned, the Judd-Ofelt parameters are obtained under the assumption that all Stark sublevels are equally populated, but, as we have shown, this may not be true at room temperature for sufficiently large level splitting within a manifold, which possibly leads to the temperature-dependence of the $\Omega_\lambda$ parameters \cite{Nhuong_2023, Ciric_2020}.

\subsection{Simulations}

The luminescence spectra can be simulated if one knows the crystal-field energy level structure and all the Stark-Stark transition strengths. The former can be obtained by diagonalizing the crystal-field Hamiltonian in the spin-orbit manifolds \cite{Hanninen_2010}. The transition strengths for single Stark-Stark transitions are scarcely reported in the literature (\textit{e.g.} [\!\!\citenum{Leavitt_1980, Burdick_1994, Gruber_2008}]). This is partially because of i) the difficulty in spectrally separating single Stark-Stark lines in the emission or absorption spectrum and ii) the simplification that comes in the Judd-Ofelt theory when one sums over all Stark-Stark line strengths: the orthogonality of the 3j-symbols makes possible to join all crystal-related parameters into the Judd-Ofelt parameters $\Omega_\lambda$, as in Eq. \eqref{eq:judd_ofelt_def} \cite{Walsh_2006}.

\begin{figure*}[h!] 
\begin{center}
	\includegraphics[width=\linewidth]{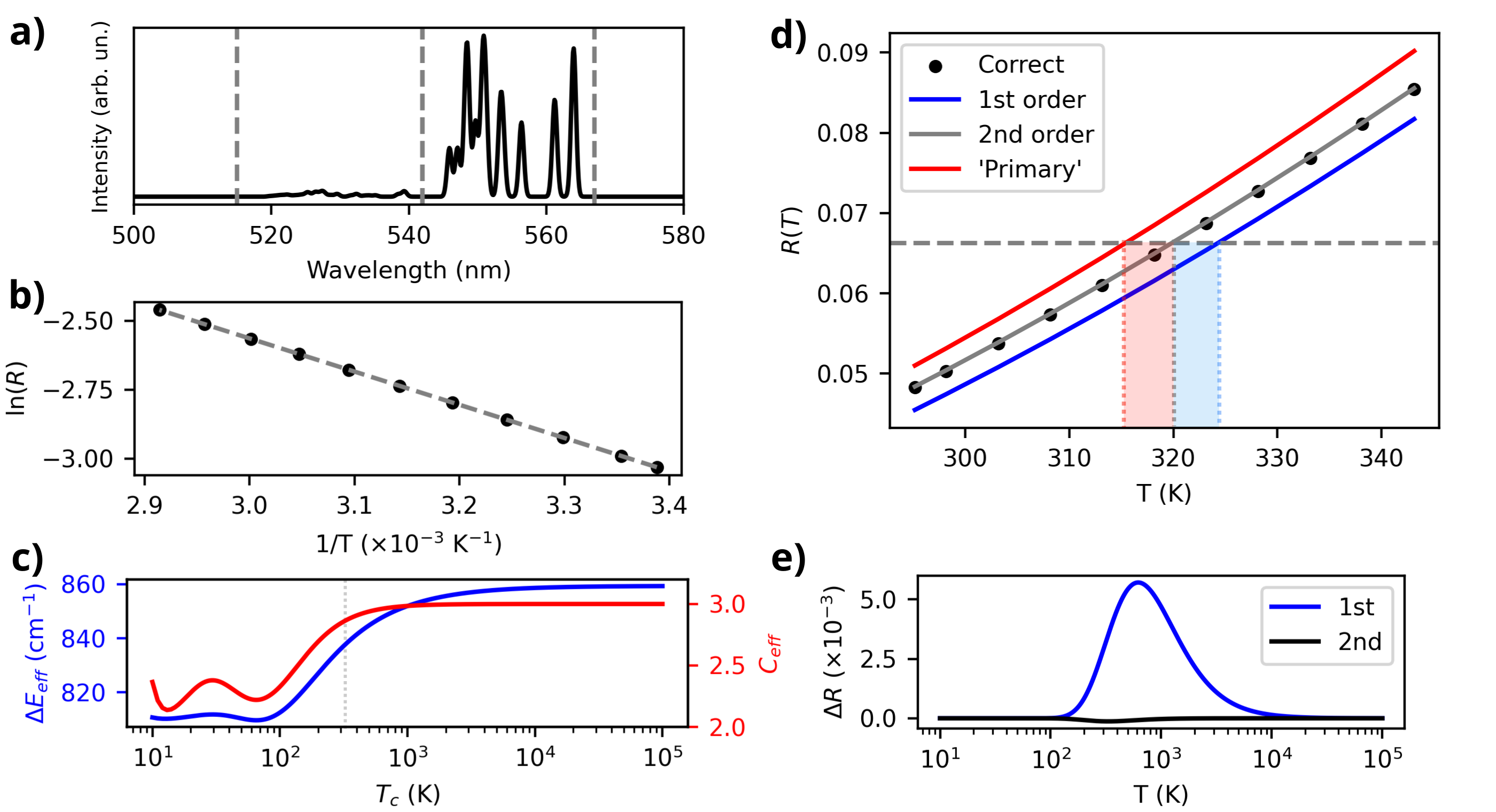}
	\caption{a) Simulated luminescence spectrum of the spectral bands $^{2}$H$_{11/2}$ $\rightarrow$ $^{4}$I$_{15/2}$ and $^{4}$S$_{3/2}$ $\rightarrow$ $^{4}$I$_{15/2}$ of Er$^{3+}$ ions in the Y$_2$O$_3$ host at \qty{295}{\kelvin}. The spectra were simulated in the idealized situation where all Stark-Stark lines have the same transition strength. b) $\ln(\text{LIR})$ as a function of the inverse temperature calculated from simulated spectra between \qty{22}{\degreeCelsius} and \qty{70}{\degreeCelsius}. c) Temperature dependence of the calibration parameters. d) Comparison between the correct temperature dependence for the simulated system, and those obtained by approximations. e) Difference between the correct temperature dependence of $R(T)$ and its first- and second-order power-series expansion.}
	\label{fig:simulations}
\end{center}
\end{figure*}

To better exemplify the role of the crystal-field splitting on the determination of the thermometric parameters, we simulated the luminescence spectrum of a Y$_2$O$_3$:Er$^{3+}$ system by using the energy level structure determined by Kisliuk \textit{et. al} \cite{Kisliuk_1964}. For simplicity, the line strengths were initially set equal for all Stark-Stark transitions between the $^2$H$_{11/2}$ and $^4$S$_{3/2}$ to the ground state, $^4$I$_{15/2}$. The population of the TCLs follows Eq. \eqref{eq:boltzmann_distribution}. The lineshape function in the wavelength space is considered as a Gaussian function  $\tilde L_{ik,jl}(\lambda) = (\sigma \sqrt{2\pi})^{-1} \exp(-(\lambda-\lambda_0)^2/(2\sigma^2))$, where  $\lambda_0$ is the central wavelength for each Stark-Stark transition. $\sigma$ is related to the linewidth of the lines in the wavelength space. We fixed $\sigma$ as \qty{0.35}{\nano\meter}, so that the FWHM of the Stark-Stark lines are $\sim$\qty{0.8}{\nano\meter}, comparable to lines observed experimentally in Y$_2$O$_3$:Er$^{3+}$ systems \cite{Pessoa_2023}. The spectrum in the energy space is obtained based on the transformation of Eq. \eqref{eq:energy_wvl_transf} \cite{Mooney_2013}. The predicted spectrum at \qty{295}{\kelvin} is shown in Fig. \ref{fig:simulations}a). It is worth emphasizing that our analysis assumes that the spectral lines of the two TCLs are well-separated in wavelength so that one can integrate the $|2\rangle \rightarrow |0\rangle$ and $|1\rangle \rightarrow |0\rangle$ bands independently. We have used a Gaussian function as the lineshape function for the sake of simplicity. A more complete description would include a Voigt profile \cite{Casabone_2018}, although more parameters are needed. Nevertheless, it does not change the results below, provided that $\int_{\lambda_1}^{\lambda_2} L_{ik,jl}(\lambda) \, d\lambda \approx 1$ for the chosen integration interval $[\lambda_1, \lambda_2]$.

We can calculate the energy barycenter of each manifold-to-manifold transition as the weighted average of the respective spectral band in the energy space, resulting in $\Delta E_\text{bary} = $ \qty{835.5}{\per\centi\meter} at \qty{295}{\kelvin}, while $\Delta E_\text{avg} = $ \qty{859.3}{\per\centi\meter} from the data from Kisliuk \textit{et at.} \cite{Kisliuk_1964}. For the thermometry characterization, we simulated spectra for different temperatures in the range of \qtyrange{22}{70}{\degreeCelsius} and used those spectra in the fitting procedure ($\ln(R)$ \textit{vs.} $T^{-1}$ curve - Eq. \eqref{eq:LIR_fitting}), as shown in Fig. \ref{fig:simulations}b). The integration interval is [515 nm, 542 nm] for the $^2$H$_{11/2}$ band, and [542 nm, 567 nm] for the $^4$S$_{3/2}$. We obtained the calibration parameters $C_\text{eff} = $ \qty{2.86}{} and $\Delta E_\text{eff} = $ \qty{837.2}{\per\centi\meter}. This shows that neither $g_2A_{20}/g_1A_{10}$ or $\Delta E_\text{bary}$ (or even $\Delta E_\text{avg}$) are adequate parameters to reproduce the experimental (in this case, simulated) data in this temperature range, even in this ideal situation. On the other hand, according to the prediction of Eq. \eqref{eq:DeltaE_eff_andC_eff}, if one performs a thermometric characterization in the temperature range of \qtyrange{22}{70}{\degreeCelsius} ($T_c=$ \qty{319.15}{\kelvin}), one obtains $C_\text{eff} = 2.86$ and $\Delta E_\text{bary} = $ \qty{837.3}{\per\centi\meter}, showing an excellent agreement with those obtained through the simulations. 

Table \ref{tab:delta_E_comparison} shows the values found in this analysis. Figure \ref{fig:simulations}c) depicts the temperature dependence of the calibration parameters by using Eq. \eqref{eq:DeltaE_eff_andC_eff}. Notice that, as predicted, $C_\text{eff}$ approaches $A_{20}g_2/A_{10}g_1 = g_2/g_1 = 3$ for high temperatures ($>$ \qty{e3}{\kelvin}), while $\Delta E_\text{eff}$ approaches $\overline{\mathrm{E}_2}_\mathcal{\mathit{w}}-\overline{\mathrm{E}_1}_\mathcal{\mathit{w}}$, which in this simplified case of all $A_{ik,0m}$ equal, is $\Delta E_\text{eff} = \Delta E_\text{bary} = \Delta E_\text{avg} =$ \qty{859.3}{\per\centi\meter} as $T \rightarrow \infty$. This is not the case when the Stark-Stark lines have different strengths. In section S1 of the Supporting Information (SI), we provide the same simulations by using a randomly generated set of Stark-Stark line strengths. For the specific simulations shown in section S1, $\Delta E_\text{eff}$ approaches \qty{864.39}{\per\centi\meter}, while $\Delta E_\text{bary}$ approaches \qty{827.53}{\per\centi\meter} as $T \rightarrow \infty$ (Fig. S2). Figure \ref{fig:simulations}c) shows an important consequence for several experiments because most luminescence thermometry studies are performed at relatively low temperatures (\qty{e2}{\kelvin} to \qty{e3}{\kelvin}), where $\Delta E_\text{eff}$ and $C_\text{eff}$ are shown to vary significantly even in this idealized model.

\begin{table}[h!]
\caption{\label{tab:delta_E_comparison}%
Comparison between different quantities related to the TCLs' energy difference. Boltzmann fitting considers an ideal Boltzmann equilibrium between the TCLs. Energies are in units of \qty{}{\per\centi\meter}.
}

\begin{tabular}{ccrc}
\hline
\multirow{2}{*}{$\Delta E_\text{avg}$} & \multirow{2}{*}{$\Delta E_\text{bary}$$^a$} & \multicolumn{2}{c}{Boltzmann fit}                                    \\
                                       &                                         & $\Delta E_\text{eff}$                & $C_\text{eff}$                \\ \hline
859.3                                  & 835.5                                   & 837.2                                & 2.86                          \\
                                       &                                         & \multicolumn{2}{c}{Prediction of Eq. \eqref{eq:DeltaE_eff_andC_eff}} \\ \cline{3-4} 
                                       &                                         & 837.3                                & 2.86                          \\ \hline
\end{tabular}
     \begin{tablenotes}
       \item [a] $^a$ At \qty{22}{\degreeCelsius}.
     \end{tablenotes} 
\end{table}

The discrepancy between $C_\text{eff}$ and $A_{20}g_2/A_{10}g_1$ and between $\Delta E_\text{eff}$ and $\Delta E_\text{bary}$ for measurements performed at room temperature shows that it is not conceptually correct to rely solely on such microscopic quantities for building primary thermometers. This is because i) the equivalence $C_\text{eff} = A_{20}g_2/A_{10}g_1$ is only valid in the infinite temperature limit; and ii) $\Delta E_\text{eff}$ may differ from $\Delta E_\text{bary}$ even at $T \rightarrow \infty$. To have an accurate Ln$^{+3}$-based primary Boltzmann luminescence thermometer using manifold-to-manifold transitions, without the aid of any external temperature measurement, it is essential to know all Stark-Stark line strengths.

\subsubsection{Implications on primary thermometer's accuracy}

As discussed in the previous sections, the current standard procedure to build a `theoretical' primary thermometer is to use the LIR's temperature dependence as in Eq. \eqref{eq:LIR_fitting}, with $C_\text{eff}$ based on Judd-Ofelt theory (Eq. \eqref{eq:judd_ofelt_def}) and $\Delta E_\text{eff}$ as the energy barycenter at room temperature. But we have demonstrated that such approach is not conceptually correct. To understand the implications of such discrepancies in thermometer's accuracy, we show in Fig. \ref{fig:simulations}d) the correct temperature dependence of $R(T)$, calculated by using Eq. \eqref{eq:LIR_maifold_manifold}, along with this theoretical `primary' approach, where the $R(T)$ curve is generated by using  Eq. \eqref{eq:LIR_fitting} with $C_\text{eff} = A_{20}g_2/A_{10}g_1 = 3$ and $\Delta E_\text{eff} = \Delta E_\text{bary} =$ \qty{835.5}{\per\centi\meter}, at \qty{297}{\kelvin}. For completeness, Fig. \ref{fig:simulations}d) also shows the first- and second-order expansion of Eq. \eqref{eq:LIR_Cumulant_generating} in terms of $\beta$. The first-order expansion is Eq. \eqref{eq:ln_LIR_cumulant}, while the second-order expansion is, explicitly,

\begin{equation}
\begin{aligned}
	\ln{R(T)} &\approx \ln{\left(\frac{g_2 \, A_{20}}{g_1 \, A_{10}} \right)} - \frac{(\overline{\mathrm{E}_2}_\mathcal{\mathit{w}}-\overline{\mathrm{E}_1}_\mathcal{\mathit{w}})}{k_{\text{B}}T} \; \\
    &+\frac{\mathrm{\sigma}_{2\mathcal{\mathit{w}}}^2 - \mathrm{\sigma}_{1\mathcal{\mathit{w}}}^2}{2(k_\text{B}T)^2} \, ,
    \end{aligned}
    \label{eq:LIR_expansion_to2ndOrder}
\end{equation}

\noindent being $\sigma_{i\mathcal{\mathit{w}}}^2 = \sum_{k}\, \mathcal{\mathit{w}}_{ik} \, \, E_{ik}^2 - \overline{\mathrm{E}_i}_\mathcal{\mathit{w}}$, as already defined. 

We can observe that the second-order approximation agrees well with the correct LIR values in this temperature range, while the first-order function and the `primary' method present high discrepancies. We calculate the LIR value corresponding to \qty{320}{\kelvin} based on the correct temperature dependency, Eq. \eqref{eq:LIR_maifold_manifold}, then we use it to compute what would be the predicted temperature for the other curves. The predicted temperature in the 1st-order approximation would be \qty{324.3}{\kelvin}, while in the 2nd-oder approximation would be \qty{319.8}{\kelvin}. For the `primary' method, we would get \qty{315.4}{\kelvin}, a difference of \qty{4.6}{\kelvin} from the correct temperature. It is worth noticing that this calculation is based in the idealized situation where all $A_{ik,0m}$ are equal. For different Stark-Stark strengths, the difference can be even higher (See section S1 of the SI, where the corresponding difference is more than \qty{20}{\kelvin}). Furthermore, Fig. \ref{fig:simulations}e) shows the difference between the correct LIR value and the values obtained by using a first- and second-order approximation functions. 

Notice that, for low-temperature range measurements, the $\ln(R)$ vs. $T^{-1}$ curve is well-described by a straight line with coefficients given by Eq. \eqref{eq:DeltaE_eff_andC_eff}. Trying to fit such a dependency with a higher-order function is not numerically stable, as the additional parameters can introduce spurious correlations and overfitting, leading to unreliable coefficient estimations and reduced predictive accuracy.

\section{Influence of the photophysical dynamics}

Besides the effects of thermalization within the thermally coupled manifolds, electronic dynamical phenomena other than phonon-mediated population exchange can also exert an influence on the LIR temperature dependence, consequently on the determination of $C_\text{eff}$ and $\Delta E_\text{eff}$ through the $\ln(R)$ \textit{vs.} $T^{-1}$ curve. It is worth emphasizing that the situation discussed in the previous sections considers no other route of population or depopulation of the TCLs, leading to an ideal Boltzmann distribution.

In Eq. \eqref{eq:phonon_rate_equation}, by including terms representing depopulation processes, $\kappa_a n_a$ and $\kappa_b n_b$, one can write

\begin{equation}
\begin{aligned} 
	\dot{n_b} &= n_0\cdot \Phi + n_a \cdot W_{a \rightarrow b}^{\text{abs}} (T) - n_b \cdot W_{b \rightarrow a}^{\text{dec}} (T) - \kappa_b n_b\\
    \dot{n_a} &= - n_a \cdot W_{a \rightarrow b}^{\text{abs}} (T) + n_b \cdot W_{b \rightarrow a}^{\text{dec}} (T) - \kappa_a n_a\\
    n_0 &= N^\prime - n_a - n_b \;,
\end{aligned} 
	\label{eq:rate_equation_plus_constant} 
\end{equation}

\noindent where $\Phi$ is the photon absorption rate from the excitation light. In steady-state one can solve it to obtain the population ratio between the TCLs as 

\begin{equation}
\begin{aligned} 
r = \frac{n_b}{n_a} = \frac{W_{a \rightarrow b}^{\text{abs}} (T)}{W_{b \rightarrow a}^{\text{dec}} (T)}+\frac{\kappa_a}{W_{b \rightarrow a}^{\text{dec}} (T)} \; .
\label{eq:adimentional_r}
\end{aligned} 
\end{equation}

\noindent The first term leads to the Boltzmann factor, as in Eq. \eqref{eq:phonon_LIR} while the influence of the second term depends on the relation between $\kappa_a$ and $W_{b \rightarrow a}^{\text{dec}} (T)$. The bigger $\kappa_a$ compared to $W_{b \rightarrow a}^{\text{dec}} (T)$, the greater the deviation from the `perfect' Boltzmann thermometer. It is insightful to investigate the effect of such a depopulation route on the relative sensitivity (Eq. \eqref{eq:def_sens_rel}), which connects directly with the calibration parameters $C_\text{eff}$ and $\Delta E_\text{eff}$ by means of Eq. \eqref{eq:DeltaE_eff_andC_eff}. Considering that the radiative decay rates $A_{ij}$ are temperature-independent in the range of interest, one can write $S_r = \frac{1}{r}\frac{\partial r}{\partial T} $, and by using the phonon-mediated transition rates as in Eq. \eqref{eq:non_rad_decay} and \eqref{eq:non_rad_absorp}, it leads to

\begin{equation}
\begin{aligned} 
S_r &= \frac{\Delta E_{ab}}{k_\text{B}T^2} \left( 1 - \frac{1}{r} \frac{\kappa_a(T)}{W_{b \rightarrow a}^{\text{dec}} (T)} (\langle n_{eff} \rangle+1)\right) \\
&+ \frac{1}{r} \frac{1}{W_{b \rightarrow a}^{\text{dec}} (T)} \frac{\partial \kappa_a}{\partial T} \;.
\label{eq:rel_sens_with_add_term}
\end{aligned} 
\end{equation}

\noindent If $\kappa_a$ is a radiative decay rate, considered to be temperature-independent, for instance, then its effect is to decrease the relative sensitivity, since $\langle n_{eff} \rangle$ is limited between 0 and 1, for zero and infinite temperature, respectively. As a result, $\Delta E_\text{eff}$ will also be reduced. For TCLs in Ln$^{3+}$ ions, $W_{b \rightarrow a}^{\text{dec}} (T)$ at room temperature is much larger than the $A_{ij}$, with the ratio $A_{ij}/W_{b \rightarrow a}^{\text{dec}} (\sim 300 \,\text{K}) \approx 10^{-5}$, depending on the system \cite{Dechao_2016}, meaning that the correction on the population due to the radiative decay can be negligible. Even if $A_{ij}$ is considered to be temperature-dependent due to vibronic coupling \cite{Suta_Meijerink_2020}, the coupling constant is very low for Ln$^{3+}$ ions, remaining a negligible correction.

\begin{figure}[h!] 
	\includegraphics[width=\linewidth]{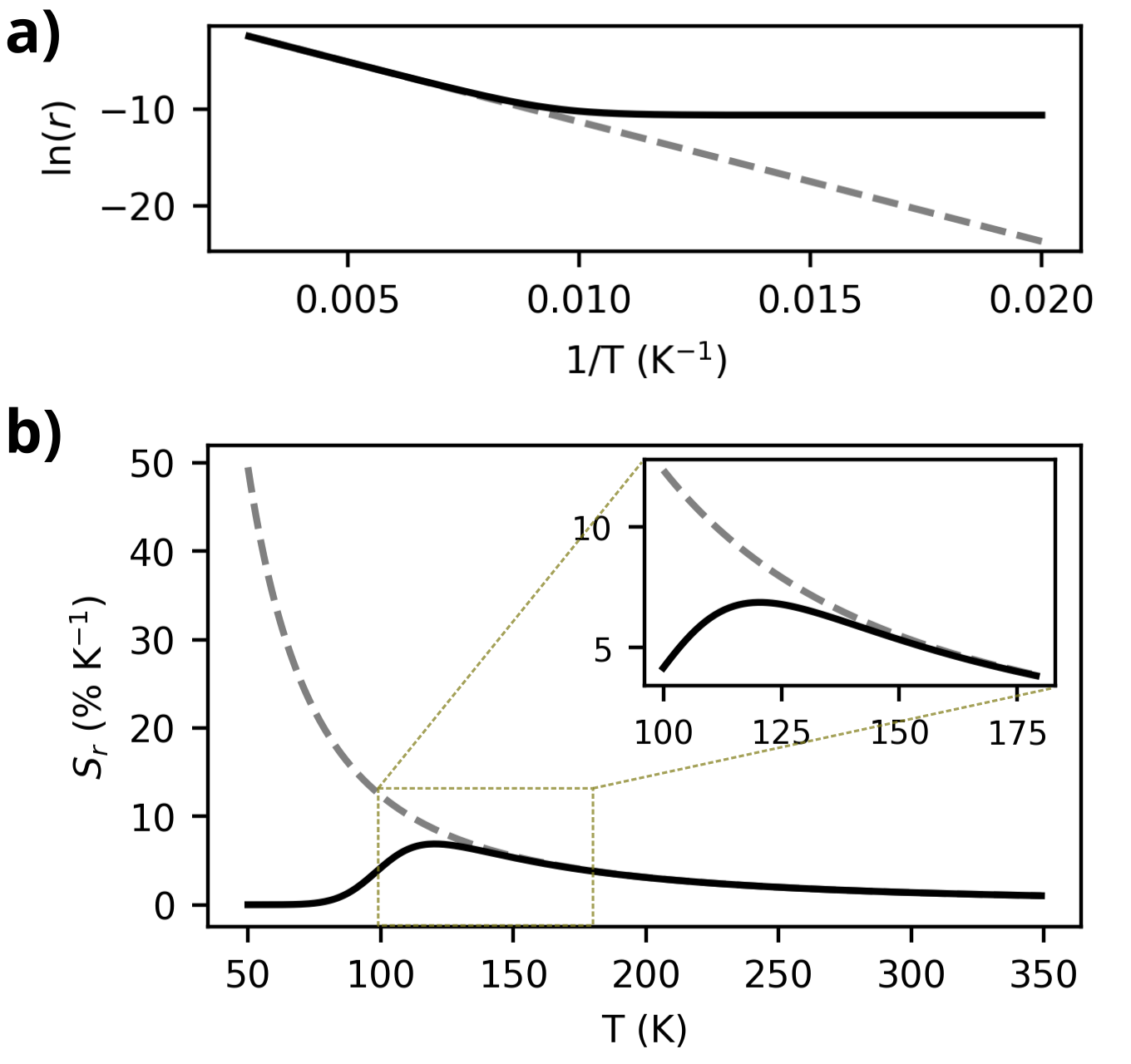}
	\caption{Behavior of a) the population ratio and b) the relative sensitivity for a Boltzmann thermometer with a temperature-independent depopulation constant (solid black curves) in comparison with a `perfect' Boltzmann thermometer (dashed gray curves). The inset in b) shows the same $S_r$ \textit{vs.} $T$ curve, but for the temperature range between \qty{100}{\kelvin} and \qty{180}{\kelvin}.}
	\label{fig:simulations_depopulation}
\end{figure}

Many dynamical processes can be modeled by incorporating them in the constant $\kappa_a$. For instance, cross-relaxation effects involving the ground state depend on a coupling constant $k^\text{CR}$ and on the populations $n_a$ and $n_0$, such that the depopulation rate is $k^\text{CR} \, n_a \, n_0$. For a suitable range of excitation powers where $n_0$ can be considered constant, then $\kappa_a = k^\text{CR}\, n_0$. As a result, the greater the cross-relaxation rate, the greater its influence on the thermometer's relative sensitivity. This is supported by experiments with Nd$^{3+}$ ions doped in a LaPO$_4$ host \cite{Suta_Miroslav_2020}. By increasing the Nd$^{3+}$ concentration, the relative sensitivity goes to zero for sufficiently low temperatures. Figure \ref{fig:simulations_depopulation} shows a) the population ratio (Eq. \ref{eq:adimentional_r}) and b) the relative sensitivity (Eq. \ref{eq:rel_sens_with_add_term}) of the thermometer described by Eqs. \eqref{eq:rate_equation_plus_constant}, where there is an additional depopulation term acting on the TCLs, in comparison with a perfect Boltzmann thermometer. The nonradiative transition rates were calculated based on Eqs. \eqref{eq:non_rad_decay} and \eqref{eq:non_rad_absorp} with the ratio $\kappa_a/W_\text{NR}^{(0)}$ assumed as \qty{e-4}{} \cite{Dechao_2016}.Also, we considered the degeneracies $g_a = 12$ and $g_b = 4$ of the TCLs of the Er$^{3+}$ ions. The energy difference was set \qty{860}{\per\centi\meter}, and the effective phonon energy is \qty{430}{\per\centi\meter}.

For high temperatures, where the phonon-mediated transitions dominate over $\kappa_a$, $\Delta E_\text{eff}$ will approximate the `ideal' Boltzmann prediction. However, at lower temperatures, Boltzmann equilibrium is lost. This behavior is observed in the LIR vs. Temperature curves of the experimental work of Suta \textit{et al.} \cite{Suta_Miroslav_2020}. Furthermore, the rate equation system (Eqs. 	\eqref{eq:rate_equation_plus_constant}) can model effects such as the influence of surrounding media on the thermometer's relative sensitivity by the coupling of the TCLs with surface ligands \cite{Galindo_2021}.


\section{Conclusions}

This study presents a comprehensive theoretical investigation of Boltzmann luminescence thermometry, shedding light on the relationship between the macroscopic calibration parameters $C_\text{eff}$ and $\Delta E_\text{eff}$ and the microscopic parameters $g_2A_{20}/g_1A_{10}$ and $\Delta E_\text{bary}$. The discrepancies between these quantities often observed among many works in the literature can be primarily attributed to two factors: i) internal thermalization among the Stark sublevels and ii) photophysical dynamics that disrupt Boltzmann equilibrium. The importance of internal thermalization stems from the comparable magnitude of the Stark splitting and $k_B T$ and at room temperature for most matrices. We show that when Stark sublevel thermalization is considered, the LIR no longer follows a simple exponential function, as commonly assumed. Instead, we derive its correct temperature dependence, making possible to obtain the difference between $\Delta E_\text{eff}$ and $\Delta E_\text{bary}$ and between $C_\text{eff}$ and $g_2A_{20}/g_1A_{10}$. Due to this difference, we show that current `primary' thermometry proposals based solely on such microscopic quantities can lead to temperature measurement errors of more than 20 K, depending on the specific system. Nevertheless, a reliable primary thermometer could be achieved if one knows all Stark-Stark line strengths for the luminescence spectrum being observed. Secondly, photophysical dynamics that disrupt Boltzmann equilibrium, influenced by experimental conditions such as excitation intensity, doping concentration, and the surrounding medium, also significantly contribute to $\Delta E_\text{eff}$ and $C_\text{eff}$. Together, these findings emphasize the need to account for both intrinsic (due to the measurement process itself) and extrinsic (due to experimental conditions) factors when developing Boltzmann thermometers.

\begin{acknowledgement}
L. de S. Menezes acknowledges the support from the Center for Nanoscience (CeNS), Ludwig Maximilians-Universit\"at M\"unchen, Germany.
\end{acknowledgement}

\begin{suppinfo}
Simulations of the thermometric characterization for random Stark-Stark line strengths are provided in the Supporting Information.
\end{suppinfo}

\vspace{1em}

\textbf{CRediT author statement}: A. R. PESSOA: Conceptualization, Formal analysis, Writing- Original draft; L. de S. MENEZES: Supervision, Project Administration, Writing- review \& editing; A. M. AMARAL: Supervision, Project Administration, Writing- review \& editing.

\onecolumn{
\bibliography{references}}

\pagebreak

\onecolumn{
\begin{figure}[h]
\begin{center}
\includegraphics[width=3.33in]{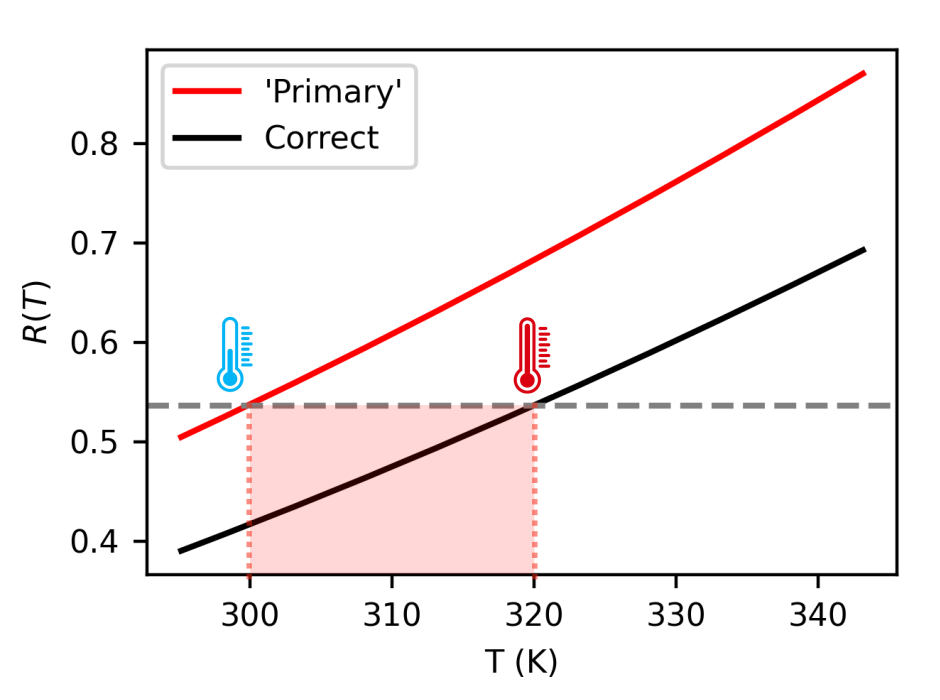}
\caption{TOC Graphic}%
\label{TOC Graphic}
\end{center}
\end{figure}
}



\section*{Supplementary material}

# Supporting Information: Addressing Discrepancies Between Theory and Experiments in Boltzmann Luminescence Thermometry with Ln$^{3+}$ Ions

Allison R. Pessoa,*,†,‡ Leonardo de S. Menezes,¶,‡ and Anderson M. Amaral‡

†Federal Institute of Education, Science and Technology of Pernambuco, Recife (50740-545), Pernambuco, Brazil

‡Department of Physics, Universidade Federal de Pernambuco, Recife (50740-540), Pernambuco, Brazil

¶Chair in Hybrid Nanosystems, Faculty of Physics, Ludwig-Maximilians-Universität München, München (80539) Bavaria, Germany

E-mail: allison.pessoa@ufpe.br

## S1 Spectrum simulations for random Stark-Stark line strengths

To exemplify the situation where the $A_{ik,0m}$ are all different, we firstly generate a set of 48 random numbers between 0 and 10 for all Stark-Stark transitions between the $^2$H$_{11/2} \to {}^4$I$_{15/2}$ manifolds of the Er$^{3+}$ ions. Another set of 16 random numbers between 0 and 1 were similarly assigned for the transitions rates of the $^4$S$_{3/2} \to {}^4$I$_{15/2}$ manifolds. The value of 10 for the first case was arbitrarily chosen so that the LIR at room temperature is similar to experimental data.[S1–S3] The generated numbers are shown in the Tables S1 and S2. The units are not important for these simulation since all quantities involving the $A_{ik,0m}$ are ratiometric. Therefore, we neglect this issue for the sake of simplicity.

Table S1: Used values for $A_{ik,0m}$, $i = 2$ ($^2H_{11/2}$ manifold), representing the transition rates for $|2\Gamma_k\rangle \to |0\Gamma_m\rangle$.

|  | $|2\Gamma_1\rangle$ | $|2\Gamma_2\rangle$ | $|2\Gamma_3\rangle$ | $|2\Gamma_4\rangle$ | $|2\Gamma_5\rangle$ | $|2\Gamma_6\rangle$ |
|---|---|---|---|---|---|---|
| $|0\Gamma_1\rangle$ | 8.59011727 | 1.02122028 | 5.3706405 | 6.62536616 | 2.06717303 | 3.65282397 |
| $|0\Gamma_2\rangle$ | 4.1135251 | 3.19410038 | 0.9185923 | 7.61588851 | 7.0140768 | 8.46816663 |
| $|0\Gamma_3\rangle$ | 5.71705033 | 1.73901931 | 4.81747164 | 9.61079451 | 1.99894644 | 8.0432751 |
| $|0\Gamma_4\rangle$ | 1.70374106 | 1.99596982 | 5.0341383 | 1.51861623 | 7.01620352 | 6.21288843 |
| $|0\Gamma_5\rangle$ | 4.61762516 | 9.95973024 | 3.62847739 | 0.83869362 | 2.17112127 | 4.88317328 |
| $|0\Gamma_6\rangle$ | 4.25211187 | 4.32476667 | 6.51482038 | 6.9084087 | 1.63331915 | 9.64024742 |
| $|0\Gamma_7\rangle$ | 4.1164009 | 9.38576801 | 9.55910253 | 2.51535831 | 2.46180075 | 4.36444428 |
| $|0\Gamma_8\rangle$ | 4.66264081 | 8.9232876 | 0.60958199 | 9.108506 | 4.86711659 | 1.62720399 |

Table S2: Used values for $A_{ik,0m}$, $i = 1$ ($^4S_{3/2}$ manifold), representing the transition rates for $|1\Gamma_k\rangle \to |0\Gamma_m\rangle$.

|  | $|1\Gamma_1\rangle$ | $|1\Gamma_2\rangle$ |
|---|---|---|
| $|0\Gamma_1\rangle$ | 0.38415509 | 0.66121677 |
| $|0\Gamma_2\rangle$ | 0.63208866 | 0.83633216 |
| $|0\Gamma_3\rangle$ | 0.97290507 | 0.57097709 |
| $|0\Gamma_4\rangle$ | 0.80402364 | 0.62924381 |
| $|0\Gamma_5\rangle$ | 0.78132722 | 0.66425061 |
| $|0\Gamma_6\rangle$ | 0.00098213 | 0.67098781 |
| $|0\Gamma_7\rangle$ | 0.94166118 | 0.18016937 |
| $|0\Gamma_8\rangle$ | 0.55806447 | 0.20453325 |

We simulated a luminescence spectrum (Fig. S1a) as described in the main manuscript, using the crystal-field energy level structure as provided in Kisliuk et al.,[S4] but setting the $A_{ik,0m}$ as those randomly generated above. Having all Stark energy levels and all line strengths, we are able to predict the correct temperature dependence of the LIR (Eq. (15) of the main text) and calculate the zeroth-, first- and second-order terms of the expansion (Eq. (29) of the main text). Using the values from the Tables S1 and S2, one gets the microscopic parameters presented in Table S3:

Table S3: Calculated microscopic values for the $Y_2O_3$: $Er^{3+}$ thermometric system based on the randomly-generated Stark-Stark line strengths.

| Microscopic quantity | Calculated value |
| --- | --- |
| $\frac{g_2 A_{20}}{g_1 A_{10}}$ | 24.82 |
| $\overline{E_2}_w - \overline{E_1}_w$ | $864.39\,\text{cm}^{-1}$ |
| $\sigma_{2w}^2 - \sigma_{1w}^2$ | $5321.27\,(\text{cm}^{-1})^2$ |

Fig. S1b) shows the $\ln(\text{LIR})$ vs. $T^{-1}$ fitting of the simulated spectra for temperatures ranging from 22 °C to 70 °C. The obtained macroscopic parameters are $C_{\text{eff}} = 23.61$ and $\Delta E_{\text{eff}} = 841.59\,\text{cm}^{-1}$, as it would be predicted by Eq. (19) of the main text. In Fig. S1c) those parameters as shown as a function of the temperature, in the low-temperature range approximation (Eq. (19)). The energy difference between the barycenters of the TCLs' luminescence bands at 295 K is calculated as $\Delta E_{\text{bary}} = 799.09\,\text{cm}^{-1}$, very different from the obtained $\Delta E_{\text{eff}}$.

With the microscopic quantities in Table S3, we can define the first-order expansion of the LIR:

$$\ln R(T) = \ln\left(\frac{g_2 A_{20}}{g_1 A_{10}}\right) - \frac{(\overline{E_2}_w - \overline{E_1}_w)}{k_B T}, \tag{S1}$$

The second-order expansion of the LIR:

$$\ln R(T) = \ln\left(\frac{g_2 A_{20}}{g_1 A_{10}}\right) - \frac{(\overline{E_2}_w - \overline{E_1}_w)}{k_B T} + \frac{\sigma_{2w}^2 - \sigma_{1w}^2}{2(k_B T)^2}, \tag{S2}$$

And the function based on the current 'primary' method:

$$\ln R(T) = \ln\left(\frac{g_2 A_{20}}{g_1 A_{10}}\right) - \frac{\Delta E_{\text{bary}}(T = 295\text{K})}{k_B T}, \tag{S3}$$

Eqs. (S1), (S2) and (S3) are plotted in Fig. S1d). The correct LIR values are calculated based on Eq. (15) of the main manuscript. Assuming that the system's temperature is 320 K, then the first-order approximation would give a measurement of 324.4 K, the second-order approximation would give a measurement of 319.9 K and the 'primary' method, commonly adopted in the literature, would result in a measurement of 299.9 K, a difference of more than 20 K. Finally, Fig. S1e) shows the

difference from the correct LIR value to that obtained with first- and second-order approximation functions.

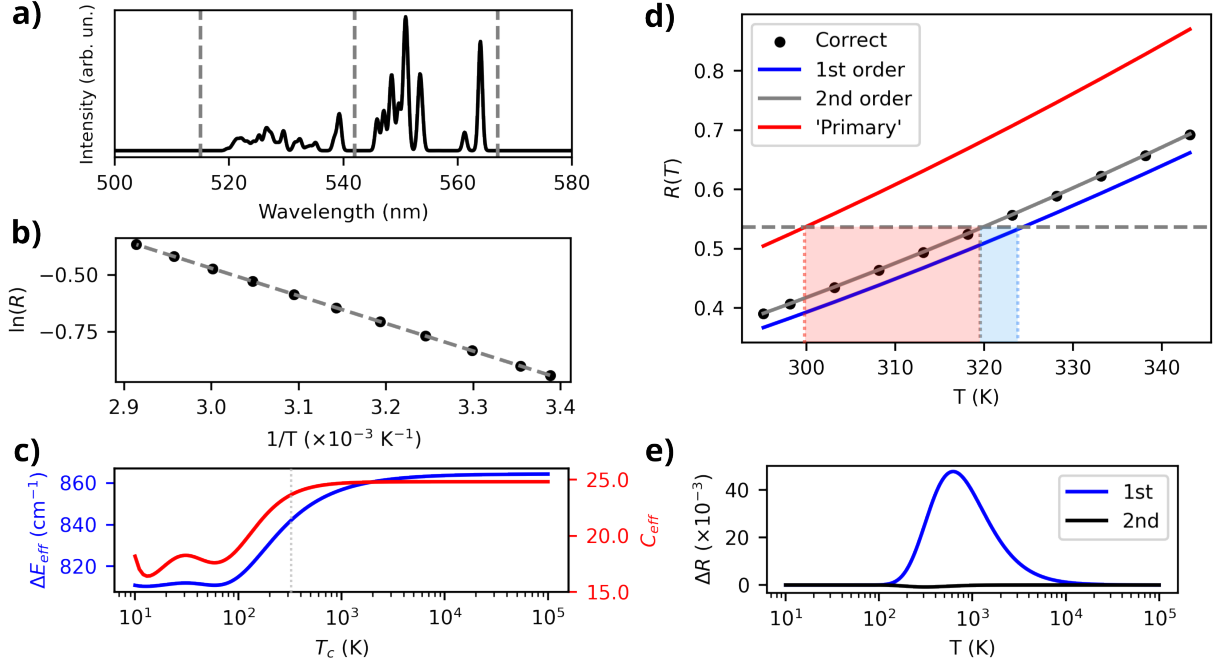

Figure S1: a) Simulated luminescence spectrum of the spectral bands $^2H_{11/2} \to {}^4I_{15/2}$ and $^4S_{3/2} \to {}^4I_{15/2}$ of $Er^{3+}$ ions in the $Y_2O_3$ host at 295 K. The spectra were simulated by using the Stark-Stark line strength presented in Tables S1 and S2. b) ln(LIR) as a function of temperature calculated from simulated spectra between 22 °C and 70 °C. c) Temperature dependence of the calibration parameters. d) Comparison between the correct temperature dependence for the simulated system, and those obtained by approximations. e) Difference between the correct temperature dependence of $R(T)$ and its first- and second-order power-series expansion.

As predicted in Eq. (27) of the main text, Fig. S2 shows $\Delta E_{\text{bary}}$ and $\Delta E_{\text{eff}}$ as a function of the temperature, for comparison, indicating that those quantities are different even in the limit of infinite temperature. For $T \to \infty$, $\Delta E_{\text{eff}}$ approaches $864.39 \, \text{cm}^{-1}$, while $\Delta E_{\text{bary}}$ approaches $827.53 \, \text{cm}^{-1}$.

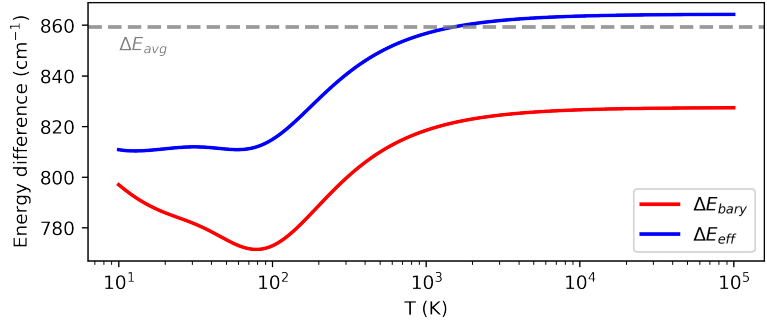

Figure S2: Comparison of $\Delta E_{\text{bary}}$ and $\Delta E_{\text{eff}}$ as a function of the temperature. The dashed horizontal line represents $\Delta E_{\text{avg}} = 859.3\,\text{cm}^{-1}$, according to the main text.



\end{document}